\documentclass[11pt]{article}
\pdfoutput=1
\usepackage{graphicx}
\usepackage{amsmath,amssymb,amsthm,amsxtra,overpic,bbm,bm,epsfig,slashed}
\usepackage{color,subfigure}
\usepackage{cite,ulem,hyperref}
\def\thefootnote{\fnsymbol{footnote}}
\allowdisplaybreaks[4]

\def\thefootnote{\fnsymbol{footnote}}

\newcommand{\bq}{\begin{eqnarray}}
\newcommand{\nq}{\end{eqnarray}}

\newcommand{\ini}{{\bf i}}
\newcommand{\fin}{{\bf f}}

\def\dx{\mathrm{d}x}
\def\dy{\mathrm{d}y}
\def\dz{\mathrm{d}z}
\def\d{\mathrm{d}}

\begin{document}

\hfill {\tt IPPP/20/38}

\begin{center}
{\Large\bf $CP$ violation in neutral lepton transition dipole moment}
\end{center} 
\vspace{0.2cm}

\begin{center}
{\bf Shyam Balaji,$^1$}\footnote{Email: \tt shyam.balaji@sydney.edu.au} ~
{\bf Maura Ramirez-Quezada$^2$}\footnote{Email: \tt maura.e.ramirez-quezada@durham.ac.uk} ~
and 
{\bf Ye-Ling Zhou$^3$}\footnote{Email: \tt ye-ling.zhou@soton.ac.uk}
\\\vspace{5mm}
{$^1$ School of Physics, The  University of Sydney, NSW 2006, Australia} \\
{$^2$ Institute for Particle Physics Phenomenology, Department of Physics, \\ Durham University, Durham DH1 3LE, United Kingdom} \\
{$^3$ School of Physics and Astronomy, University of Southampton,\\
SO17 1BJ Southampton, United Kingdom } 
\\
\end{center}

\vspace{1.5cm} 

\begin{abstract} 

The $CP$ violation in the neutrino transition electromagnetic dipole moment is discussed in the context of the Standard Model with an arbitrary number of right-handed singlet neutrinos. A full one-loop calculation of the neutrino electromagnetic form factors is performed in the Feynman gauge. A non-zero $CP$ asymmetry is generated by a required threshold condition for the neutrino masses along with non-vanishing $CP$ violating phases in the lepton flavour mixing matrix. We follow the paradiagm of $CP$ violation in neutrino oscillations to parametrise the flavour mixing contribution into a series of Jarlskog-like parameters. This formalism is then applied to a minimal seesaw model with two heavy right-handed neutrinos denoted $N_1$ and $N_2$. We observe that the $CP$ asymmetries for decays into light neutrinos $N\to \nu\gamma$ are extremely suppressed, maximally around $10^{-17}$. However the $CP$ asymmetry for $N_2 \to N_1 \gamma$ can reach of order unity. Even if the Dirac $CP$ phase $\delta$ is the only source of $CP$ violation, a large $CP$ asymmetry around $10^{-5}$-$10^{-3}$ is comfortably achieved. 

\end{abstract}
\begin{flushleft}
\hspace{0.8cm} PACS number(s): \\
\hspace{0.8cm} Keywords: $CP$ violation, neutrino dipole moment, radiative decay, seesaw mechanism
\end{flushleft}

\def\thefootnote{\arabic{footnote}}
\setcounter{footnote}{0}

\newpage

\section{Introduction}

Since the discovery of neutrino oscillations \cite{Fukuda:1998mi, Fukuda:2001nj, Ahmad:2001an, Ahmad:2002jz}, 
it has been well understood that neutrinos have tiny masses and that their flavour eigenstates are different from, but merely superpositions of their mass eigenstates. 
The mismatch between the flavour and mass basis is described by lepton flavour mixing. The most important lepton flavour question mixing remaining is whether $CP$ is violated. A large $CP$ violation is supported by the combined analysis of current accelerator neutrino oscillation data \cite{Abe:2018wpn} in the appearance channel of neutrino oscillations \cite{Abe:2017uxa, Adamson:2017gxd}. The next-generation large-scale neutrino experiments DUNE and T2HK are projected to observe $CP$ violation in the near future \cite{Abi:2018alz,Abe:2015zbg,Abe:2016ero}.

On the theoretical side, the origin of finite but tiny neutrino
masses is still unknown.
The canonical seesaw mechanism \cite{Minkowski:1977sc,Yanagida:1979as, GellMann:1980vs,Glashow:1979nm,Mohapatra:1979ia,Schechter:1980gr} and its numerous variations are proposed to solve this problem. The basic idea is that the small masses of left-handed neutrinos are attributed to the existence of
much heavier right-handed Majorana neutrinos. 
In this elegant picture the flavour states are dominantly superpositions of massless left-handed neutrinos but also, to a smaller degree, their heavy right-handed counterparts. 
The minimal seesaw model \cite{Frampton:2002qc} is a simplified version of the canonical seesaw mechanism with only two right-handed neutrinos, which has been studied in depth \cite{Xing:2020ald}.  
The seesaw mechanism induces new sources of $CP$ violation in the heavy neutrino sector, providing the so-called leptogenesis, as one of the most popular mechanisms to explain the observed matter-antimatter asymmetry in our Universe \cite{Fukugita:1986hr}. 

Neutrinos are usually considered as electrically neutral particles which do not participate in tree-level electromagnetic interactions. However, they may have electric and magnetic dipole moments  appearing at loop level. 
The study of the neutrino dipole moment dates back four decades \cite{Shrock:1974nd,Petcov:1976ff,Marciano:1977wx,Lee:1977tib}. 
In the Standard Model (SM), weak charged current interactions contribute in the loops and induce non-zero dipole moment for neutrinos \cite{Fujikawa:1980yx,Schechter:1981hw,Pal:1981rm,Schechter:1981cv,Shrock:1982sc,Nieves:1981zt,Kayser:1982br,Bilenky:1987ty}, see also in \cite{Dvornikov:2003js, Dvornikov:2004sj, Xing:2012gd}. 
A {\it transition} dipole moment between two different neutrino mass
eigenstates can trigger a heavier neutrino radiatively decaying to a lighter neutrino through the release of a photon. In fact, if neutrinos are Majorana particles, the property that Majorana
fermions are their own antiparticles implies that neutrinos have only
a transitional component to their dipole moment \cite{Xing:2011zza}.

In various studies of the neutrino dipole moment in the literature, $CP$ symmetry is always considered as an explicit symmetry for the relevant mass regions of neutrinos. However, a $CP$ violating dipole moment has many interesting phenomenological applications. It may contribute to leptogenesis to explain the observed baryon-antibaryon asymmetry in our Universe \cite{Bell:2008fm}. It also provides a source of a circular polarisation of photons in the sky for a suitable range of neutrino masses,  \cite{Boehm:2017nrl}.  
In Ref. \cite{Balaji:2019fxd}, the general conditions required to generate $CP$ violation in the dipole moment was elucidated as well as the $CP$ asymmetry based on a widely studied Yukawa interaction. The latter was applied to both left- and right-handed neutrino radiative decay scenarios as well as searches for dark matter via direct detection and collider signatures. 

This work will focus on discussing $CP$ violation in the neutrino dipole moment with right-handed neutrinos. We will provide the one-loop  calculation of the $CP$ asymmetry of the neutrino transition dipole moment in full detail in the framework of the SM with the addition of $SU(2)_L$-singlet right-handed neutrinos. In Section~\ref{sec:framework}, we review the model-independent neutrino dipole moment written in terms of form factors producing $CP$ violation. Section~\ref{sec:CP} contributes to a comprehensive analytical one-loop calculation of form factors. Finally, a numerical scan of the $CP$ asymmetry with inputs of current neutrino oscillation data is performed in Section \ref{sec:numerical}. We summarise our results in Section~\ref{sec:conclusion}. 

\section{Neutrino electromagnetic dipole moment with $CP$ violation \label{sec:framework}}

In this section we give a brief review of the framework for $CP$ violation in neutrino radiative decays. We refer to our former paper Ref. \cite{Balaji:2019fxd} for the detailed derivation.  
Discussions in Section \ref{subsec:Dirac} assumes neutrinos are Dirac particles. The extension to Majorana neutrinos will be given in Section \ref{subsec:Majorana}.

\subsection{Form factors for Dirac neutrino\label{subsec:Dirac}}

Assuming the decaying fermion is a Dirac particle, amplitudes for the processes $\nu_\ini \to \nu_\fin \gamma_+$ and $\nu_\ini \to \nu_\fin  \gamma_-$,  with respect to the photon polarisation $+$ and $-$ are given by
\begin{eqnarray}
i \mathcal{M} (\nu_\ini \to \nu_\fin \gamma_{\pm})  = i \bar{u}(p_\fin) \Gamma_{\fin \ini }^\mu(q^2) u(p_\ini) \varepsilon^*_{\pm,\mu}(q) \,, \label{eq:decay_amplitude}
\end{eqnarray}
where $u(p_\ini)$ and $u(p_\fin)$ are spinors for the initial $\nu_\ini$ and final $\nu_\fin$ state neutrinos respectively, and the photon momentum $q=p_\ini - p_\fin$. The vertex function $\Gamma_{\fin \ini }^\mu(q^2)$ can in general be decomposed into four terms, electric charge, magnetic dipole moment, electric dipole moment and the anapole form factors \cite{Nieves:1981zt, Shrock:1982sc, Kayser:1982br, Kayser:1984ge}. Without introducing a source for the electric charge, the neutrino will remain electrically neutral forever. By requiring the photon to be on-shell $q^2=0$ and choosing the Lorenz gauge $q\cdot \varepsilon_p = 0$, the anapole does not contribute to $\Gamma^\mu_{\fin \ini }$. Therefore, the vertex function is simplified to \cite{Nieves:1981zt, Shrock:1982sc, Kayser:1982br, Kayser:1984ge}
\begin{eqnarray} \label{eq:dipole}
\Gamma^\mu_{\fin \ini }(q^2=0) &=& -f^{\rm{M}}_{\fin \ini}  (i \sigma^{\mu \nu} q_\nu) + f^{\rm{E}}_{\fin \ini}  (i \sigma^{\mu \nu} q_\nu \gamma_5) \,, 
\end{eqnarray}
where $f^{\rm{E}}_{\fin \ini} $ and $f^{\rm{M}}_{\fin \ini}$ are the electric and magnetic transition dipole moments of $\nu_\ini \to \nu_\fin \gamma$ respectively. 
It is helpful to rewrite it in the chiral form 
\begin{eqnarray} \label{eq:form_factor_nu}
\Gamma^\mu_{\fin \ini } (0) &=& i \sigma^{\mu \nu} q_\nu [f^\text{L}_{\fin \ini } P_{\text{L}} + f^\text{R}_{\fin \ini } P_{\text{R}}] \,, 
\end{eqnarray}
where $f^{\text{L,R}}_{\fin \ini} = -f^{\rm{M}}_{\fin \ini} \pm i f^{\rm{E}}_{\fin \ini} $ and the chiral projection operators are defined as $P_{\text{L,R}} = \frac{1}{2}(1 \mp \gamma_5)$ \cite{Balaji:2019fxd}. 
The amplitudes $\mathcal{M}(\nu_{\ini} \to \nu_{\fin}  \gamma_{\pm})$ are directly correlated with the coefficients as \cite{Balaji:2019fxd}
\begin{eqnarray} \label{eq:amplitudes_1}
\mathcal{M}(\nu_{\ini} \to \nu_{\fin}  \gamma_+) = \sqrt{2} f^{\rm L}_{\fin \ini } (m_\ini^2 - m_\fin^2) \,, \quad
\mathcal{M}(\nu_{\ini} \to \nu_{\fin}  \gamma_-) = -\sqrt{2} f^{\rm R}_{\fin \ini } (m_\ini^2 - m_\fin^2) \,.  
\end{eqnarray}
With the above justification, decay widths for $\nu_\ini \to \nu_\fin \gamma_{\pm}$, after averaging over the spin for the initial neutrino, can be written in a simple form
\begin{eqnarray}
\Gamma(\nu_\ini \to \nu_\fin \gamma_+)
= {\cal A} | f_{\fin \ini}^{\rm L} |^2\,,\quad
\Gamma(\nu_\ini \to \nu_\fin \gamma_-)
= {\cal A} | f_{\fin \ini}^{\rm R} |^2\,,
\end{eqnarray}
with ${\cal A} = (m_\ini^2-m_\fin^2)^3/(16\pi m^3_\ini)$. 
The total radiative decay width $\Gamma(\nu_\ini \to \nu_\fin \gamma)$ is obtained by summing the decay widths for $\nu_\ini \to \nu_\fin \gamma_+$ and $\nu_\ini \to \nu_\fin \gamma_-$. 

For antineutrinos, amplitudes for $\bar{\nu}_\ini \to \bar{\nu}_\fin  \gamma_+$ and $\bar{\nu}_\ini \to \bar{\nu}_\fin  \gamma_-$ are given by 
\begin{eqnarray} \label{eq:decay_amplitude_nubar}
i \mathcal{M} (\bar{\nu}_\ini \to \bar{\nu}_\fin  \gamma_{\pm}) 
&=& i \bar{v}(p_\ini) \bar{\Gamma}_{\ini \fin}^\mu(q^2) v(p_\fin) \varepsilon^*_{\pm,\mu}(q) \,, 
\end{eqnarray}
respectively, where $v(p_\ini)$ and $v(p_\fin)$ are antineutrino spinors. 
The vertex function $\bar{\Gamma}^\mu_{\ini \fin}$ when the photon is on-shell is consequently written in a similar form as shown in Eq.~\eqref{eq:form_factor_nu}, 
\begin{eqnarray}
\bar{\Gamma}^\mu_{\ini \fin} (0) &=& i \sigma^{\mu \nu} q_\nu [\bar{f}^\text{L}_{\ini \fin} P_{\text{L}} + \bar{f}^\text{R}_{\ini \fin} P_{\text{R}}] \,.
\end{eqnarray}
Where $CPT$ invariance ensures
$ \bar{f}^\text{L}_{\ini \fin} = - f^\text{L}_{ \ini \fin}$, and
$ \bar{f}^\text{R}_{\ini \fin} = - f^\text{R}_{ \ini \fin}$ \cite{Giunti:2014ixa}. 
Hence, amplitudes $\mathcal{M}(\bar{\nu}_{\ini} \to \bar{\nu}_{\fin}  \gamma_+)$  are simplified to \cite{Balaji:2019fxd}
\begin{eqnarray} \label{eq:amplitudes_2}
\mathcal{M}(\bar{\nu}_{\ini} \to \bar{\nu}_{\fin}  \gamma_+) = \sqrt{2} f^{\rm L}_{\ini \fin} (m_\ini^2 - m_\fin^2) \,, \quad
\mathcal{M}(\bar{\nu}_{\ini} \to \bar{\nu}_{\fin}  \gamma_-) = -\sqrt{2} f^{\rm R}_{\ini \fin} (m_\ini^2 - m_\fin^2) \,.
\end{eqnarray}
The antineutrino decay widths are then given by
$\Gamma(\bar{\nu}_\ini \to \bar{\nu}_\fin \gamma_+)
={\cal A} \left| f_{\ini \fin}^{\rm L} \right|^2$ and 
$\Gamma(\bar{\nu}_\ini \to \bar{\nu}_\fin \gamma_-)
={\cal A} \left| f_{\ini \fin}^{\rm R} \right|^2$. 

In \cite{Balaji:2019fxd}, we have defined a set of $CP$ asymmetries between neutrino radiative decay and antineutrino radiative decay. In terms of ratios specifying photon polarisations, we may write 
\begin{eqnarray}
\Delta_{CP,+} = \frac{\Gamma(\nu_\ini \to \nu_\fin \gamma_+) - 
\Gamma(\bar{\nu}_\ini \to \bar{\nu}_\fin \gamma_-)}
{\Gamma(\nu_\ini \to \nu_\fin \gamma) + 
\Gamma(\bar{\nu}_\ini \to \bar{\nu}_\fin \gamma)} \,, \quad
\Delta_{CP,-} = \frac{\Gamma(\nu_\ini \to \nu_\fin \gamma_-) - 
\Gamma(\bar{\nu}_\ini \to \bar{\nu}_\fin \gamma_+)}
{\Gamma(\nu_\ini \to \nu_\fin \gamma) + 
\Gamma(\bar{\nu}_\ini \to \bar{\nu}_\fin \gamma)}\,,
\end{eqnarray}
which can further be simplified to
\begin{eqnarray} \label{eq:CP_Dirac}
\Delta_{CP,+} = \frac{|f^{\text{L}}_{\fin \ini }|^2 - |f^{\text{R}}_{\ini \fin}|^2}
{|f^{\text{L}}_{\fin \ini }|^2 + |f^{\text{R}}_{\fin \ini }|^2 + |f^{\text{R}}_{\ini \fin}|^2 + |f^{\text{L}}_{\ini \fin}|^2} \,,\quad
\Delta_{CP,-} = \frac{|f^{\text{R}}_{\fin \ini }|^2 - |f^{\text{L}}_{\ini \fin}|^2}
{|f^{\text{L}}_{\fin \ini }|^2 + |f^{\text{R}}_{\fin \ini }|^2 + |f^{\text{R}}_{\ini \fin}|^2 + |f^{\text{L}}_{\ini \fin}|^2} \,.
\end{eqnarray}
In the case of $CP$ conservation,  $f^{\rm L,R}_{ \ini \fin} = [f^{\rm R,L}_{ \fin \ini}]^*$, we arrive at vanishing $CP$ asymmetries $\Delta_{CP,+} = \Delta_{CP,-} = 0$.

\subsection{Form factors for Majorana neutrinos\label{subsec:Majorana}}

We now extend the discussion to Majorana neutrinos. 
The Majorana field satisfies $\nu = C \overline{\nu}^T$, where $C$ is the charge-conjugation matrix. Compared with the Dirac field which contains independent left-handed and right-handed components $\nu_{\rm L} \equiv P_{\rm L}\nu$ and $\nu_{\rm R} \equiv P_{\rm R} \nu$, the Majorana field enforces the right-handed component to be the charge conjugation of the left-handed component, i.e., $P_{\rm R} \nu = C\overline{\nu_{\rm L}}^T$, leading to the quantisation in the form $\nu \sim a u(p) e^{-i p\cdot x} + a^\dag v(p) e^{i p\cdot x}$. Taking this into account and applying the parametrisation in Eqs.~\eqref{eq:decay_amplitude} and \eqref{eq:decay_amplitude_nubar}, the amplitude for $\nu_\ini \to \nu_\fin  \gamma_{\pm}$ is proven to be   
\begin{eqnarray}
i \mathcal{M}^{\rm M}(\nu_\ini \to \nu_\fin  \gamma_{\pm}) = 
i \bar{u}(p_\fin) \Gamma_{\fin \ini }^\mu(q^2) u(p_\ini) \varepsilon^*_{\pm,\mu}(q) - 
i \bar{v}(p_\ini) \Gamma_{\ini \fin}^\mu(q^2) v(p_\fin) \varepsilon^*_{\pm,\mu}(q)
\end{eqnarray}
in the Majorana case \cite{Giunti:2014ixa}. 
It can be explained as the sum of amplitudes of the Dirac neutrino radiative decay and antineutrino radiative decay channels, 
i.e., $i \mathcal{M}^{\rm M}(\nu_\ini \to \nu_\fin  \gamma_{\pm}) = i \mathcal{M} (\nu_\ini \to \nu_\fin  \gamma_{\pm}) + i \mathcal{M} (\bar{\nu}_\ini \to \bar{\nu}_\fin  \gamma_{\pm}) 
$.
Taking the explicit formulas for the amplitudes given in Eq.~\eqref{eq:amplitudes_1} and Eq.~\eqref{eq:amplitudes_2}, we obtain results with definite spins in the initial and final states as 
\begin{eqnarray} \label{eq:amplitude_Majorana}
\hspace{-5mm}
\mathcal{M}^{\rm M}(\nu_{\ini} \to \nu_{\fin}  \gamma_+) = +
\sqrt{2} [ f^{\text{L}}_{\fin \ini } - f^{\text{L}}_{\ini \fin} ] (m_\ini^2 - m_\fin^2) \,,~
\mathcal{M}^{\rm M}(\nu_{\ini} \to \nu_{\fin}  \gamma_-) = -
\sqrt{2} [ f^{\text{R}}_{\fin \ini } - f^{\text{R}}_{\ini \fin} ] (m_\ini^2 - m_\fin^2) \,. 
\end{eqnarray}
The decay widths are given by $\Gamma^{\rm M}(\nu_\ini \to \nu_\fin  \gamma_+) = {\cal A} |f^{\text{L}}_{\fin \ini } - f^{\text{L}}_{\ini \fin}|^2$ and  $\Gamma^{\rm M}(\nu_\ini \to \nu_\fin  \gamma_-) = {\cal A} |f^{\text{R}}_{\fin \ini } - f^{\text{R}}_{\ini \fin}|^2$. 

For Majorana fermions, the $CP$ violation is identical to that obtained from $P$-violation alone i.e. 
the $CP$ asymmetry is essentially the same as the asymmetry between the two polarised photons. Hence, we have 
\begin{eqnarray} \label{eq:CP_Majorana}
\Delta_{CP,+}^{\rm M} = - \Delta_{CP,-}^{\rm M} = \frac{\Gamma^{\rm M}(\nu_\ini \to \nu_\fin \gamma_{+})  
- \Gamma^{\rm M}(\nu_\ini \to \nu_\fin \gamma_{-}) }
{\Gamma^{\rm M}(\nu_\ini \to \nu_\fin+ \gamma)}
=
\frac{|f^{\text{L}}_{\fin \ini } - f^{\text{L}}_{\ini \fin}|^2 - |f^{\text{R}}_{\fin \ini } - f^{\text{R}}_{\ini \fin}|^2}
{|f^{\text{L}}_{\fin \ini } - f^{\text{L}}_{\ini \fin}|^2 + |f^{\text{R}}_{\fin \ini } - f^{\text{R}}_{\ini \fin}|^2} \,.
\end{eqnarray}
For simplicity, we make the assignment $\Delta_{CP}\equiv \Delta_{CP,+}^{\rm M}$ for use in the following phenomenological discussions.

\section{$CP$ violating form factors induced by charged-current interactions \label{sec:CP}}

We present below, the one-loop calculation of neutrino radiative decay $\nu_\ini \to \nu_\fin \gamma$ for massive neutrinos with the existence of $CP$ violation. We work in the framework of the SM extended with an arbitrary number of $SU(2)_L$-singlet right-handed neutrinos in the Feynman gauge.
The crucial operator for the charged-current interaction is
\begin{eqnarray} \label{eq:cc}
\mathcal{L}_{\rm c.c.} = \sum_{\alpha, m}  \frac{g}{\sqrt{2}} {\cal U}_{\alpha m} \; \bar{\ell}_\alpha \gamma^\mu  P_{\rm L} \nu_m W^{-}_{\mu} + {\rm h.c.}\,,
\end{eqnarray}
where $g$ is the electroweak (EW) gauge coupling constant, $\alpha$ is an index that represents charged lepton flavours $\alpha=e,\mu,\tau$ and $m$ is an index that represents the neutrino mass eigenstates. In particular, $\nu_m = \nu_1, \nu_2, \nu_3$ represent three light neutrino mass eigenstates and $\nu_m = N_1, N_2, \dots$ representing heavy neutrino mass eigenstates. The matrix ${\cal U}_{\alpha m}$ denotes the lepton flavour mixing accounting for heavy neutrino mass eigenstates. 

\begin{figure}[t!]
\centering
\includegraphics[width=.6\textwidth]{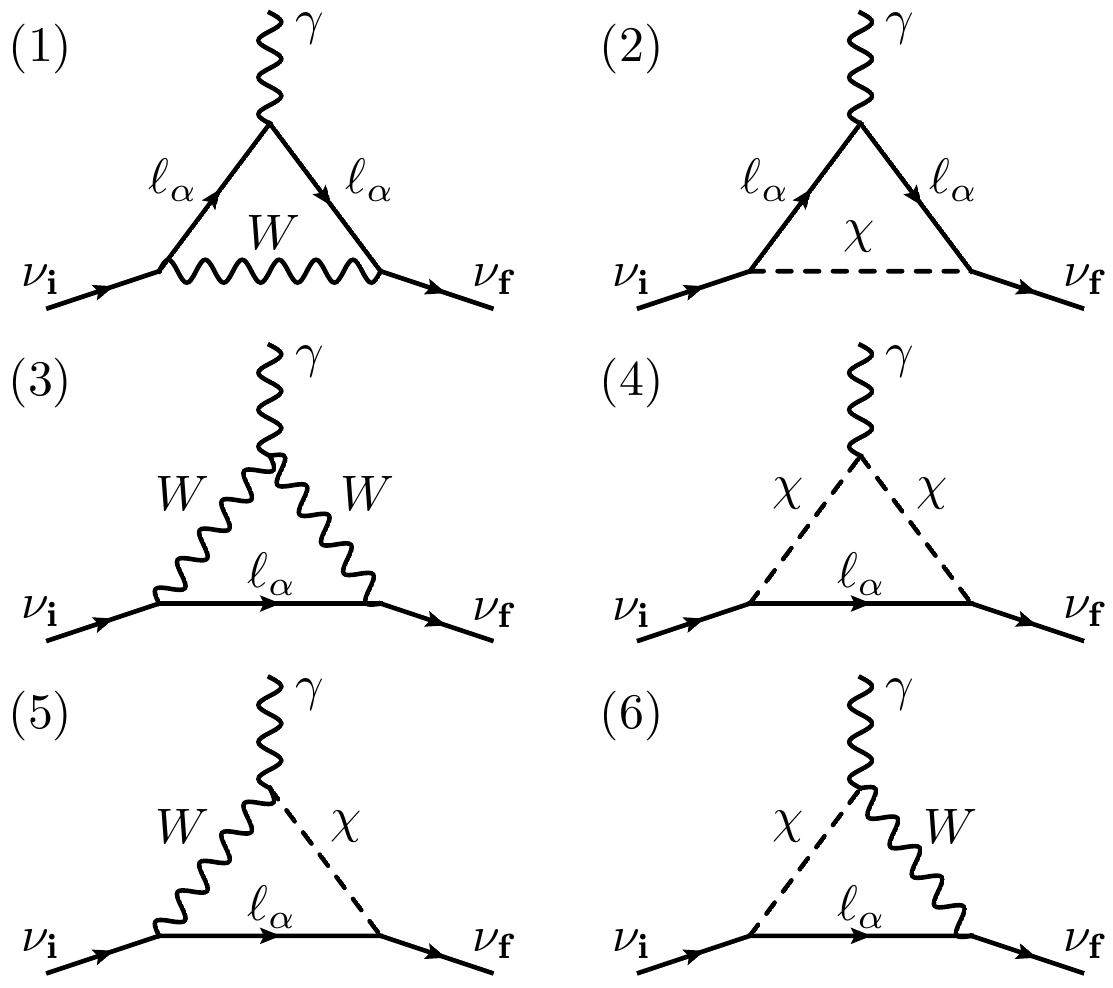}
  \caption{All Feynman diagrams contributing to the neutrino electromagnetic transition dipole moment, where $\chi$ is the charged Goldstone boson. }\label{fig:loop_SM_ints}
\end{figure}   

The one-loop Feynman diagrams for the radiative decay via the SM charged current interaction are shown in Fig.~\ref{fig:loop_SM_ints}. 
The vertex functions of each
proper vertex diagram in Fig. 1 is given by
\begin{eqnarray}
\Gamma^{\mu,(1)}_{\fin\ini,\alpha} &=& i{\frac{eg^{2}}{2}}
  \mathcal{U}_{\alpha \ini} \mathcal{U}^*_{\alpha \fin} \int \frac{\d^4 p}{(2\pi)^4}
  {\frac{\gamma_{\nu} P_{\rm L}
  (\slashed{p}_\fin-\slashed{p}+m_{\alpha})\gamma^{\mu}
  (\slashed{p}_\ini-\slashed{p}+m_{\alpha})\gamma^{\nu}P_{\rm L}}
  {[(p_\fin-p)^{2}-m^2_\alpha]
  [(p_\ini-p)^{2}-m^2_\alpha][p^{2}-m_W^2]}}\;,\nonumber\\
\Gamma^{\mu,(2)}_{\fin\ini,\alpha} &=& i{\frac{eg^{2}}{2}}
  \mathcal{U}_{\alpha \ini} \mathcal{U}^*_{\alpha \fin}\int \frac{\d^4 p}{(2\pi)^4}
  {\frac{(m_\fin P_{\rm L}-m_{\alpha}P_{\rm R})
  (\slashed{p}_\fin-\slashed{p}+m_{\alpha})\gamma^{\mu}
  (\slashed{p}_\ini-\slashed{p}+m_{\alpha})
  (m_{\alpha}P_{\rm L}-m_\ini P_{\rm R})}
  {m_W^2[(p_\fin-p)^{2}-m^2_\alpha]
  [(p_\ini-p)^{2}-m^2_\alpha][p^{2}-m_W^2]}}\;,\nonumber\\
\Gamma^{\mu,(3)}_{\fin\ini,\alpha} &=& i\frac{e g^2}{2}
  \mathcal{U}_{\alpha \ini} \mathcal{U}^*_{\alpha \fin} \int \frac{\d^4 p}{(2\pi)^4}
\frac{ \gamma_\rho P_{\rm L} (\slashed{p}+m_\alpha) \gamma_\nu P_{\rm L} V^{\mu\nu\rho} } { \large[ (p_\fin-p)^2 -
  m_W^2 \large] \large[ (p_\ini-p)^2 - m_W^2 \large] \large[ p^2
  - m^2_\alpha \large] }\;,\nonumber\\
\Gamma^{\mu,(4)}_{\fin\ini,\alpha} &=&
  i{\frac{eg^{2}}{2}}
  \mathcal{U}_{\alpha \ini} \mathcal{U}^*_{\alpha \fin} \int \frac{\d^4 p}{(2\pi)^4}
  {\frac{(2p-p_\ini-p_\fin)^{\mu}(m_\fin P_{\rm L}-m_{\alpha}P_{\rm R})
  (\slashed{p}+m_{\alpha})
  (m_{\alpha}P_{\rm L}-m_\ini P_{\rm R})}
  {m_W^2[(p_\fin-p)^{2}-m_W^2]
  [(p_\ini-p)^{2}-m_W^2][p^{2}-m^2_\alpha]}}\;,\nonumber\\
\Gamma^{\mu,(5)}_{\fin\ini,\alpha} &=&
  i{\frac{eg^{2}}{2}}
  \mathcal{U}_{\alpha \ini} \mathcal{U}^*_{\alpha \fin} \int \frac{\d^4 p}{(2\pi)^4}
  {\frac{\gamma^{\mu}P_{\rm L} (\slashed{p} + m_{\alpha})
  (m_{\alpha}P_{\rm L}-m_\ini P_{\rm R})}
  {[(p_\fin-p)^{2}-m_W^2]
  [(p_\ini-p)^{2}-m_W^2][p^{2}-m^2_\alpha]}}\;,\nonumber\\
\Gamma^{\mu,(6)}_{\fin\ini,\alpha} &=&
  i{\frac{eg^{2}}{2}}
  \mathcal{U}_{\alpha \ini} \mathcal{U}^*_{\alpha \fin} \int \frac{\d^4 p}{(2\pi)^4}
  {\frac{(m_{\alpha}P_{\rm R} - m_\fin P_{\rm L})
  (\slashed{p} + m_{\alpha})\gamma^{\mu}P_{\rm L}}
  {[(p_\fin-p)^{2}-m_W^2]
  [(p_\ini-p)^{2}-m_W^2][p^{2}-m^2_\alpha]}}\;,
\end{eqnarray}
where
\begin{eqnarray}
V^{\mu\nu\rho} &=& g^{\mu\nu}(2p_\ini-p-p_\fin)^\rho +
  g^{\rho\mu}(2p_\fin-p-p_\ini)^\nu +
  g^{\nu\rho}(2p-p_\ini-p_\fin)^\mu \,.
\end{eqnarray}

The non-vanishing $CP$ asymmetry requires two conditions. Namely, a $CP$ violating contribution from coefficients of tree-level vertices and an imaginary part coming purely from loop kinematics \cite{Balaji:2019fxd}. In the present work, the first condition is satisfied by the complex phases in the lepton flavour mixing matrix ${\cal U}$ and will be discussed in more detail in subsequent sections. Here, we first contend with the second condition by completing the loop calculation and deriving its imaginary part analytically.

We follow the standard procedure to integrate the loop momenta with the help of the Feynman parametrisation. Then, we apply the Gordon decomposition taking chirality into consideration, and factorise dipole moment terms with coefficients as 
\begin{eqnarray}\label{eq:effective_vertex1}
&\Gamma_{\fin \ini, \alpha}^{\mu,({\rm k})}&= \frac{e g^2}{4(4\pi)^2} \mathcal{U}_{\alpha \ini} \mathcal{U}^*_{\alpha \fin}
i \sigma^{\mu\nu}q_\nu \int_0^1\dx\dy\dz\,\delta(x+y+z-1)\, {\cal P}^{(\rm k)} \,,
\end{eqnarray}
where 
\begin{eqnarray}\label{eq:effective_vertex2}
&{\cal P}^{(1)}&= \frac{-2x (x+z) m_\ini P_\text{R} - 2x (x+y) m_\fin P_\text{L}}
{\Delta_{\alpha W}(x,y,z)}\,, \nonumber\\
&{\cal P}^{(2)}&= \frac{[x z m_\fin^2 -((1 - x)^2 +xz) m_\alpha^2] m_\ini P_\text{R} + [xy m_\ini^2 - ((1-x)^2 + xy) m_\alpha^2] m_\fin P_\text{L}}
{m_W^2 \Delta_{\alpha W}(x,y,z)}\,, \nonumber\\
&{\cal P}^{(3)}&= \frac{[(1-2x)z-2(1-x)^2] m_\ini P_\text{R} + [(1-2x)y-2(1-x)^2] m_\fin P_\text{L}}
{\Delta_{W \alpha}(x,y,z)}\,, \nonumber\\
&{\cal P}^{(4)}&= \frac{[xz m_\fin^2 - x (x+z) m_\alpha^2] m_\ini P_\text{R} + [xy m_\ini^2 - x (x+y) m_\alpha^2] m_\fin P_\text{L}}
{m_W^2 \Delta_{W \alpha}(x,y,z)}\,, \nonumber\\
&{\cal P}^{(5)}&= \frac{-z m_\ini P_\text{R}} {\Delta_{W \alpha}(x,y,z)}\,, \nonumber\\
&{\cal P}^{(6)}&= \frac{-y m_\fin P_\text{L}} {\Delta_{W \alpha}(x,y,z)}\,, 
\end{eqnarray}
and
 \begin{eqnarray} \label{eq:Deltaxyz}
\Delta_{W \alpha}(x,y,z) &=& m_W^2(1-x)+x m_\alpha^2-x (y m_\ini^2 + z m_\fin^2) \,, \nonumber\\
\Delta_{\alpha W}(x,y,z) &=& m_\alpha^2(1-x)+x m_W^2-x (y m_\ini^2 + z m_\fin^2) \,. 
\end{eqnarray} 

Eq.~\eqref{eq:effective_vertex1} can be further simplified to  
\begin{eqnarray}\label{eq:effective_vertex3}
&\Gamma_{\fin \ini, \alpha}^{\mu,({\rm k})}&= \frac{eG_\text{F}}{4\sqrt{2}\pi^2} \mathcal{U}_{\alpha \ini} \mathcal{U}^*_{\alpha \fin}
i \sigma^{\mu\nu}q_\nu ({\cal F}_{\fin \ini, \alpha}  m_\ini P_\text{R}+ {\cal F}_{\ini \fin, \alpha} m_\fin P_\text{L}) \,.
\end{eqnarray}
Here, $ {\cal F}$ is derived from the sum of the integrals ${\cal P}^{(k)}$
\begin{eqnarray}\label{eq:F_integral}
&&{\cal F}_{\fin \ini, \alpha} =  \int_0^1 \dx \left\{ \frac{\left(m_\ini^2-m_\alpha^2-2 m_W^2\right) \left(m_\alpha^2+m_\fin^2 x^2\right)+m^4_{\fin\ini,\alpha} x}{\left(m_\ini^2-m_\fin^2\right)^2 x} \log \left(\frac{m_\alpha^2+\left(m_W^2-m_\alpha^2-m_\ini^2\right) x+m_\ini^2 x^2}{m_\alpha^2+\left(m_W^2-m_\alpha^2-m_\fin^2\right) x+m_\fin^2 x^2}\right) \right. \nonumber\\
&&+\left. \frac{\left(m_\ini^2-m_\alpha^2-2 m_W^2\right) \left(m_\alpha^2+m_\fin^2 (1-x)^2\right)+m^4_{\fin\ini,\alpha} (1-x)}{\left(m_\ini^2-m_\fin^2\right)^2 x} 
\log \left(\frac{m_W^2+\left(m_\alpha^2-m_W^2-m_\ini^2\right) x+m_\ini^2 x^2}{m_W^2 + \left(m_\alpha^2-m_W^2-m_\fin^2\right) x+m_\fin^2 x^2}\right) \right\} \nonumber\\
&&+\frac{m_\fin^2-m_\alpha^2-2 m_W^2}{m_\ini^2-m_\fin^2} \,, 
\end{eqnarray}
where we define $m^4_{\fin\ini,\alpha} = - (m_\ini^2-m_\alpha^2-m_W^2)(m_\fin^2+m_\alpha^2-2m_W^2) + 2 m_\alpha^2 m_W^2$, and
${\cal F}_{\ini \fin, \alpha}$ is obtained by exchanging $m_\ini$ and $m_\fin$. Therefore, we obtain the coefficients $f^{\text{L}}_{\fin \ini}$, $f^{\text{L}}_{\ini \fin}$, $f^{\text{R}}_{\fin \ini}$ and $f^{\text{R}}_{\ini \fin}$ as
\begin{eqnarray}
&&f^{\text{L}}_{\fin \ini} = \frac{eG_\text{F}}{4\sqrt{2}\pi^2} \mathcal{U}_{\alpha \ini} \mathcal{U}^*_{\alpha \fin}
 {\cal F}_{\ini \fin, \alpha} m_\fin\,,\quad
 f^{\text{R}}_{\fin \ini} = \frac{eG_\text{F}}{4\sqrt{2}\pi^2} \mathcal{U}_{\alpha \ini} \mathcal{U}^*_{\alpha \fin}
 {\cal F}_{\fin \ini, \alpha} m_\ini\,,\nonumber\\
&&f^{\text{L}}_{\ini \fin} = \frac{eG_\text{F}}{4\sqrt{2}\pi^2} \mathcal{U}_{\alpha \fin} \mathcal{U}^*_{\alpha \ini}
 {\cal F}_{\fin \ini, \alpha} m_\ini\,,\quad
 f^{\text{R}}_{\ini \fin} = \frac{eG_\text{F}}{4\sqrt{2}\pi^2} \mathcal{U}_{\alpha \fin} \mathcal{U}^*_{\alpha \ini}
 {\cal F}_{\ini \fin, \alpha} m_\fin\,.
\end{eqnarray}

The integrals ${\cal F}_{\fin \ini, \alpha}$ and ${\cal F}_{\ini \fin, \alpha}$ in Eq.~\eqref{eq:F_integral} can be further simplified when the limit of small neutrino masses, i.e., $m_{\ini}^2, m_{\fin}^2 \ll m_\alpha^2,m_W^2$ is considered. In this case, the logarithm terms can be expanded in a series of $m_\ini^2$ and $m_\fin^2$, and after a straightforward calculation, we prove that both ${\cal F}_{\fin \ini, \alpha}$ and ${\cal F}_{\ini \fin, \alpha}$ are identical to $F(m_\alpha^2/m_W^2)$, where 
\begin{eqnarray} \label{eq:dipole moment_SM}
F(a) =
\frac{3}{4}\left(\frac{2-a}{1-a}-\frac{2
a}{(1-a)^2}-\frac{2 a^2\log
a}{(1-a)^3}\right)
\end{eqnarray}  
which is a well known result for the loop factor obtained in the studies of neutrino dipole moments and radiative decays \cite{Pal:1981rm,Shrock:1982sc}. 

We now outline how to obtain non-zero imaginary parts for ${\cal F}_{\fin \ini, \alpha}$ and ${\cal F}_{\ini \fin, \alpha}$ when neutrinos have large masses. 
They include integral terms of the form $\int _0^1 \dx f (x) \log g (x)$, where $g(x)$ is not always positive in the domain $(0,1)$. Instead, one can prove that there is an interval $(x_1,x_2) \subset (0,1)$ where $g(x)<0$ is satisfied, and $x_1$ and $x_2$ are solutions of $g(x)=0$. The real and imaginary parts in the integral can then be split into 
\begin{eqnarray}
\int _0^1 \dx f (x) \log g (x)=\int _0^1 \dx f(x) \log|g(x)|+ i\pi \int_{x_1}^{x_2} \dx f(x) \,.
\end{eqnarray}
The imaginary part of $\int_{x_1}^{x_2}\dx f (x)$ can then be analytical obtained. 
In this way, we derive the analytical expression for the imaginary part of ${\cal F}_{\fin \ini, \alpha}$ as
\begin{eqnarray}
\hspace{-5mm}
{\rm Im}({\cal F}_{\fin \ini, \alpha}) =\pi \vartheta(m_\ini - m_W - m_\alpha) \left\{
\frac{m_\ini^2-m_\alpha^2-2 m_W^2}{\left(m_\ini^2-m_\fin^2\right)^2} 
\left[-\mu_\ini^2 \frac{m_\fin^2}{m_\ini^2} + m_\alpha^2 \log \left(\frac{m_\ini^2+m_\alpha^2-m_W^2+\mu_\ini^2}{m_\ini^2+m_\alpha^2-m_W^2-\mu_\ini^2}\right)\right] \right. &&\nonumber\\
\left.+ \frac{\left(2 m_\ini^2-m_\fin^2-m_\alpha^2-2 m_W^2\right) m_W^2}{\left(m_\ini^2-m_\fin^2\right)^2} \log \left(\frac{m_\ini^2-m_\alpha^2+m_W^2+\mu_\ini^2}{m_\ini^2-m_\alpha^2+m_W^2-\mu_\ini^2}\right)\right\} &&\nonumber\\ 
 +  \pi \vartheta(m_\fin - m_W - m_\alpha) \left\{ 
 - \frac{m_\ini^2-m_\alpha^2-2 m_W^2}{\left(m_\ini^2-m_\fin^2\right)^2}
\left[-\mu_\fin^2+m_\alpha^2 \log \left(\frac{m_\fin^2+m_\alpha^2-m_W^2+\mu_\fin^2}{m_\fin^2+m_\alpha^2-m_W^2-\mu_\fin^2}\right)\right] \right.&& \nonumber\\
+\left.
\frac{\left(2 m_\ini^2-m_\fin^2-m_\alpha^2-2 m_W^2\right)m_W^2}{\left(m_\ini^2-m_\fin^2\right)^2} 
\log \left(\frac{m_\fin^2-m_\alpha^2+m_W^2+\mu_\fin^2}{m_\fin^2-m_\alpha^2+m_W^2-\mu_\fin^2}\right)
 \right\}&\!\!,&\label{eq:ImF}
\end{eqnarray}
where $\vartheta(x)$ is the Heaviside step function, and
\begin{eqnarray}
\mu_\ini^2 &=& \sqrt{m_\ini^4+m_\alpha^4+m_W^4-2m_\ini^2 m_\alpha^2 - 2 m_\ini^2 m_W^2 - 2 m_\alpha^2 m_W^2} \,, \nonumber\\
\mu_\fin^2 &=& \sqrt{m_\fin^4+m_\alpha^4+m_W^4-2m_\fin^2 m_\alpha^2 - 2 m_\fin^2 m_W^2 - 2 m_\alpha^2 m_W^2} \,.
\end{eqnarray}
Again, ${\rm Im}({\cal F}_{\ini \fin, \alpha})$ is obtained from ${\rm Im}({\cal F}_{\fin \ini, \alpha})$ by exchanging $m_\ini$ and $m_\fin$. Some comments on the imaginary part of ${\cal F}_{\fin \ini, \alpha}$ are
\begin{itemize} 
\item In order to generate a non-zero imaginary part in the loop integration, a threshold condition for the initial neutrino mass is required. That is $m_\ini > m_W+ m_\alpha$, namely, initial neutrino mass larger than the sum of the $W$-boson mass and the charged lepton mass. This is consistent with optical theorem as discussed in Ref. \cite{Balaji:2019fxd}.

\item Taking the charged lepton flavour to be the electron, $\alpha =e$, the threshold condition for initial neutrino masses is simplified to $m_\ini > m_W+ m_e \approx m_W$. 

\item There is a second contribution to the imaginary part of ${\cal F}_{\fin \ini, \alpha}$ if the neutrino in the final state satisfies the threshold condition, $m_\fin > m_W+ m_\alpha$. Due to the sign difference, it partly cancels with the first contribution.  

\end{itemize}

With the above results, we are now able to obtain the most general result for $CP$ asymmetries in neutrino radiative decays. For Dirac neutrinos, recall Eq.~\eqref{eq:CP_Dirac}. We derive the $CP$ asymmetry between $\nu_\ini \to \nu_\fin \gamma_+$ and $\bar{\nu}_\ini \to \bar{\nu}_\fin \gamma_-$ and between $\nu_\ini \to \nu_\fin \gamma_-$ and $\bar{\nu}_\ini \to \bar{\nu}_\fin \gamma_+$ as
\begin{eqnarray}
\Delta^{\rm D}_{CP, +} &=& \frac{ - \sum_{\alpha,\beta}{\cal J}_{\alpha\beta}^{\ini \fin}
 {\rm Im}({\cal F}_{\ini \fin, \alpha} {\cal F}_{\ini \fin, \beta}^*) m_\fin^2 
 }{
 \sum_{\alpha,\beta}
{\cal R}_{\alpha\beta}^{\ini \fin}
\left[ {\rm Re}({\cal F}_{\fin \ini, \alpha} {\cal F}_{\fin \ini, \beta}^*) m_\ini^2 + {\rm Re}({\cal F}_{\ini \fin, \alpha} {\cal F}_{\ini \fin, \beta}^*) m_\fin^2 \right]
}\,, \nonumber\\
\Delta^{\rm D}_{CP, -} &=& \frac{ - \sum_{\alpha,\beta}{\cal J}_{\alpha\beta}^{\ini \fin}
{\rm Im}({\cal F}_{\fin \ini, \alpha} {\cal F}_{\fin \ini, \beta}^*) m_\ini^2 
 }{
 \sum_{\alpha,\beta}
{\cal R}_{\alpha\beta}^{\ini \fin}
\left[ {\rm Re}({\cal F}_{\fin \ini, \alpha} {\cal F}_{\fin \ini, \beta}^*) m_\ini^2 + {\rm Re}({\cal F}_{\ini \fin, \alpha} {\cal F}_{\ini \fin, \beta}^*) m_\fin^2 \right]
} \,,
\end{eqnarray}
where $\alpha,\beta$ run for charged lepton flavours $e,\mu,\tau$ and
\begin{eqnarray} \label{eq:J_and_R}
{\cal J}_{\alpha\beta}^{\ini \fin} = {\rm Im} ({\cal U}_{\alpha \ini} {\cal U}_{\alpha \fin}^* {\cal U}_{\beta \ini}^* {\cal U}_{\beta \fin}) \,, &&
{\cal R}_{\alpha\beta}^{\ini \fin} = {\rm Re} ({\cal U}_{\alpha \ini} {\cal U}_{\alpha \fin}^* {\cal U}_{\beta \ini}^* {\cal U}_{\beta \fin}) \,.
\end{eqnarray}
We now outline the contribution of coefficients to the tree-level vertices. We have introduced a set of Jarlskog-like parameters ${\cal J}_{\alpha\beta}^{\ini \fin}$ to describe the $CP$ violation from the vertex contribution. This parametrisation follows the famous definition of the Jarlskog invariant used to describe $CP$ violation in neutrino oscillations \cite{Jarlskog:1985ht, Wu:1985ea}. The Jarlskog-like parameters are invariant under any phase rotation of charged leptons and neutrinos. If the Jarlskog-like parameters vanish, no $CP$ violation is generated in the neutrino transition dipole moment. 

For Majorana neutrinos, the relevant $CP$ asymmetries, via Eq.~\eqref{eq:CP_Majorana}, are given by 
\begin{eqnarray} \label{eq:CP_Majorana_2}
\Delta^{\rm M}_{CP, +} &=& - \Delta^{\rm M}_{CP, -} \nonumber\\
&=& \frac{\sum_{\alpha,\beta}\,{\cal J}_{\alpha\beta}^{\ini \fin}
\left[ {\rm Im}({\cal F}_{\fin \ini, \alpha} {\cal F}_{\fin \ini, \beta}^*) m_\ini^2 - {\rm Im}({\cal F}_{\ini \fin, \alpha} {\cal F}_{\ini \fin, \beta}^*) m_\fin^2 \right]
 - 2 {\cal V}_{\alpha\beta}^{\ini \fin} {\rm Im}({\cal F}_{\fin \ini, \alpha} {\cal F}_{\ini \fin, \beta}^*) m_\ini m_\fin}{\sum_{\alpha,\beta}
{\cal R}_{\alpha\beta}^{\ini \fin}
\left[ {\rm Re}({\cal F}_{\fin \ini, \alpha} {\cal F}_{\fin \ini, \beta}^*) m_\ini^2 + {\rm Re}({\cal F}_{\ini \fin, \alpha} {\cal F}_{\ini \fin, \beta}^*) m_\fin^2 \right]
 - 2 {\cal C}_{\alpha\beta}^{\ini \fin} {\rm Re}({\cal F}_{\fin \ini, \alpha} {\cal F}_{\ini \fin, \beta}^*) m_\ini m_\fin} \,,
\end{eqnarray}
where 
\begin{eqnarray} \label{eq:V_and_C}
{\cal V}_{\alpha\beta}^{\ini \fin} = {\rm Im} ({\cal U}_{\alpha \ini} {\cal U}_{\alpha \fin}^* {\cal U}_{\beta \ini} {\cal U}_{\beta \fin}^*) \,, &&
{\cal C}_{\alpha\beta}^{\ini \fin} = {\rm Re} ({\cal U}_{\alpha \ini} {\cal U}_{\alpha \fin}^* {\cal U}_{\beta \ini} {\cal U}_{\beta \fin}^*) \,.
\end{eqnarray}
${\cal V}_{\alpha\beta}^{\ini \fin}$ is another type of Jarlskog-like parameters which appears only for Majorana neutrinos. It was first defined in the study of neutrino-antineutrino oscillations in the context of only three light neutrinos \cite{Xing:2013woa}. They are invariant under phase rotations for charged lepton but not for neutrinos.

\section{$CP$ violation in heavy neutrino radiative decays \label{sec:numerical}}

In the rest of this paper, we will discuss the $CP$ violating radiative decay in the seesaw model, where the tiny masses for left-handed neutrinos are generated due to the suppression of heavy right-handed neutrinos. We recall that the notation $\Delta_{CP} = \Delta_{CP,+}^{\rm M}$ for Majorana neutrinos is used.

We consider the minimal seesaw model where only two copies of right-handed neutrinos are introduced \cite{Frampton:2002qc}. This is the minimal number required to generate two non-zero mass square differences i.e. $\Delta m_{21}^2 \equiv m_2^2 - m_1^2$ and $\Delta m_{31}^2 \equiv m_3^2 - m_1^2$. 
We denote two right-handed neutrino mass eigenstates as $N_I$ for $I=1,2$, with masses $M_1< M_2$. 
The following discussion is straightforwardly generalised to a canonical seesaw model with three right-handed neutrinos. Including more copies of right-handed neutrinos just increases the number of free model parameters. 

The minimal seesaw model predicts one massless neutrino $m_1=0$ in the normal mass ordering ($m_1<m_2<m_3$) and $m_3=0$ in the inverted mass ordering ($m_3<m_1<m_2$) schemes. In this section, we will only consider the normal mass ordering as we don't expect the inverted mass ordering to make a significant difference. Moreover, the inverted ordering is slightly disfavoured ($\Delta \chi^2 =6.2$) by the current neutrino oscillation global fit data \cite{Esteban:2018azc}. 
We take the best fit (in the $3\sigma$ ranges) of mass square differences in the normal ordering scheme \cite{Esteban:2018azc}, this is
\begin{eqnarray} \label{eq:mass_data}
m_2 &=& \sqrt{\Delta m^2_{21}} = 8.60\; (8.24 \to 8.95)~{\rm meV}\,, \nonumber\\
m_3 &=& \sqrt{\Delta m^2_{31}} = 50.2\; (49.3 \to 51.2)~{\rm meV}\,.
\end{eqnarray}
We recall once again the lepton charged-current interaction in Eq.~\eqref{eq:cc}. The three light neutrino mixing is represented by the first $3\times 3$ submatrix of ${\cal U}$, i.e., ${\cal U}_{\alpha i}$ for $\alpha=e,\mu,\tau$ and $i=1,2,3$. In the case of negligible non-unitary effect, ${\cal U}_{\alpha i}$ is parametrised as 
\begin{eqnarray}
U \equiv \left(
\begin{array}{ccc}
 c_{12} c_{13} & c_{13} s_{12} & s_{13} e^{-i \delta} \\
 -c_{23} s_{12}-c_{12} s_{13} s_{23} e^{i \delta} & c_{12} c_{23}-s_{12} s_{13} s_{23} e^{i \delta} & c_{13} s_{23} \\
 s_{12} s_{23}-c_{12} c_{23} s_{13} e^{i \delta} & -c_{12} s_{23}-c_{23} s_{12} s_{13} e^{i \delta} & c_{13} c_{23} \\
\end{array}
\right)  \left(
\begin{array}{ccc}
e^{i\rho} & 0 & 0 \\
0 & e^{i\sigma} & 0 \\
0 & 0 & 1 \\
\end{array}
\right)
\,,
\end{eqnarray}
where $c_{ij} = \cos \theta_{ij}$, $s_{ij} = \sin \theta_{ij}$, $\theta_{ij}$ (for $ij = 12, 13, 23$) are three mixing angles, $\delta$ is the Dirac-type $CP$ violating phase and $\rho$ and $\sigma$ are two Majorana-type $CP$ violating phases. $U$ is a $3 \times 3$ unitary matrix, $U^\dag U = U U^\dag = \mathbf{1}_{3 \times 3}$. The three mixing angles and the Dirac $CP$ violating phase for normal mass ordering are measured to be
\begin{eqnarray}
\label{eq:PMNS_data}
\theta_{13} &=& 8.61^\circ\;\;\;\;\, (8.22^\circ \to 8.99^\circ)\,, \nonumber\\
\theta_{12} &=& 33.82^\circ\; (31.61^\circ \to 36.27^\circ)\,, \nonumber\\
\theta_{23} &=& 48.3^\circ\;\;\;\;\, (40.8^\circ \to 51.3^\circ)\,, \nonumber\\
\delta &=& 222^\circ \;\;\;\; \;\;\, (141^\circ \to 370^\circ)
\end{eqnarray}
at the best fit (in the $3\sigma$ ranges) \cite{Esteban:2018azc}. As we work in the minimal seesaw model where the lightest neutrino mass $m_1 = 0$ is massless, $\rho$ is unphysical and will not be considered below. We are left with two $CP$ violating phases $\delta$ and $\sigma$ from the mixing of light neutrinos.

Accounting for the non-unitary effect, namely, the fraction of heavy neutrinos contributing to the flavour mixing ${\cal U}_{\alpha (I+3)}$, which we denote as $R_{\alpha I}$ from now on. 
${\cal U}_{\alpha i}$ is only approximately equal to $U_{\alpha i}$, ${\cal U}_{\alpha i} = U_{\alpha i} + {\cal O}(RR^\dag)$. $RR^\dag$ is constrained to be maximally at milli-level \cite{Fernandez-Martinez:2016lgt,Coutinho:2019aiy}. Therefore, ${\cal U}_{\alpha i} \approx U_{\alpha i}$ is still a  very good approximation.

The charged-current interaction for leptons in the mass eigenstates is now written as
\begin{eqnarray}
\mathcal{L}_{\rm c.c.} = \sum_{\alpha=e,\mu,\tau} \frac{g}{\sqrt{2}} \; \bar{\ell}_\alpha \gamma^\mu  P_{\rm L} \Big( \sum_{i=1,2,3} U_{\alpha i} \nu_i  + \sum_{I=1,2} R_{\alpha I} N_I \Big)W^{-}_{\mu} + {\cal O}(RR^\dag)
+ {\rm h.c.}\,.
\end{eqnarray}
We use the Casas-Ibarra parametrisation \cite{Casas:2001sr} to express $R$ in the form
\begin{eqnarray}
R_{\alpha I} = \sum_{i=1,2} U_{\alpha i} \Omega_{iI} \sqrt{\frac{m_{i+1}}{M_I}} \,.
\end{eqnarray} 
Here, $\Omega$ is a $2\times 2$ complex orthogonal matrix satisfying $\Omega^T \Omega = \Omega \Omega^T = \mathbf{1}$.\footnote{In the case of three copies of right-handed neutrinos, $\Omega$ is a $3\times 3$ matrix, this leads to each entry in $R_{\alpha I}$ for $I=1,2,3$ to be expressed as $$R_{\alpha I} = \sum_{i=1,2,3} U_{\alpha i} \Omega_{iI} \sqrt{\frac{m_{i}}{M_I}}\,.$$} 
We parametrise it as
\begin{eqnarray}
\Omega = \left(
\begin{array}{cc}
\cos \omega & \sin \omega \\
- \zeta \sin \omega~ & \zeta \cos \omega
\end{array}
\right)\,,\label{eq:Omega_param}
\end{eqnarray}
where $\omega$ is a complex parameter and $\zeta = \pm 1$. The two possible values of $\zeta$ correspond to two distinct branches of $\Omega$ \cite{Antusch:2011nz,DiBari:2019zcc}. 
The Yukawa coupling $Y$ between lepton doublets and right-handed neutrinos are directly connected with $R$ via $Y_{\alpha I} = R_{\alpha I} M_I/v_H$ \cite{Ibarra:2003up}. 

In the whole model, three $CP$ violating parameters are induced, $\delta$, $\sigma$ and ${\rm Im}[\omega]$, if $\delta=0$, $\sigma=0$ or $\pi/2$ and ${\rm Im}[\omega]=0$, no $CP$ violation can be generated.

The $CP$ violation in the neutrino transition dipole moment can be checked by the study of the $CP$ asymmetry of neutrino radiative decay. There are three channels of interest, $\nu_i \to \nu_j \gamma$, $N_I \to \nu_i \gamma$ and $N_2\to N_1\gamma$. For the first channel, since the light neutrinos have masses much lighter than the $W$ boson, no $CP$ violation can be generated. 
The $CP$ asymmetry for $N_I \to \nu_i \gamma$ is non-zero if $N_I$ has a mass $M_I > m_W+m_e \approx m_W$. Note that in this case, masses of three light neutrinos $\nu_i$ for $i=1,2,3$ are negligible and photons released in the relevant three channels are indistinguishable, so we sum these channels together and calculate the overall $CP$ asymmetry  [cf. Eq.~\eqref{eq:CP_Majorana_2}]
\begin{eqnarray}
\Delta_{CP} (N_I \to \nu\gamma)
&=& \frac{\sum_i \sum_{\alpha,\beta}\,{\cal J}_{\alpha\beta}^{(I+3) i}
 {\rm Im}({\cal F}_{i (I+3), \alpha} {\cal F}_{i (I+3), \beta}^*) }{\sum_i \sum_{\alpha,\beta}
{\cal R}_{\alpha\beta}^{(I+3) i}
 {\rm Re}({\cal F}_{i (I+3), \alpha} {\cal F}_{i (I+3), \beta}^*) } \,.
\end{eqnarray}
This parameter is tiny, numerically confirmed to be maximally $\lesssim 10^{-17}$. The reason why it is so small can be understood as follows. Since $m_i$ is negligible, ${\cal F}_{i(I+3),\alpha} = {\cal F}_{1(I+3),\alpha}$, and $\Delta_{CP} (N_I \to \nu\gamma) \propto \sum_i \sum_{\alpha,\beta}\,{\cal J}_{\alpha\beta}^{(I+3) i} = \sum_i \sum_{\alpha,\beta}\, {\rm Im} ({\cal U}_{\alpha (I+3)} {\cal U}_{\alpha i}^* {\cal U}_{\beta (I+3)}^* {\cal U}_{\beta i}) \approx \sum_i \sum_{\alpha}\, {\rm Im} ({\cal U}_{\alpha (I+3)} {\cal U}_{\alpha (I+3)}^*) = 0$. 

Finally, we focus on the $CP$ asymmetry in $N_2\to N_1\gamma$, which is given by 
\begin{eqnarray}
\Delta_{CP} (N_2 \to N_1 \gamma) &=&
\frac{\sum_{\alpha,\beta}\,{\cal J}_{\alpha\beta}^{5 4}
\left[ {\rm Im}({\cal F}_{4 5, \alpha} {\cal F}_{4 5, \beta}^*) M_2^2 - {\rm Im}({\cal F}_{5 4, \alpha} {\cal F}_{5 4, \beta}^*) M_1^2 \right]
 - 2 {\cal V}_{\alpha\beta}^{5 4} {\rm Im}({\cal F}_{4 5, \alpha} {\cal F}_{5 4, \beta}^*) M_2 M_1}{\sum_{\alpha,\beta}
{\cal R}_{\alpha\beta}^{5 4}
\left[ {\rm Re}({\cal F}_{4 5, \alpha} {\cal F}_{4 5, \beta}^*) M_2^2 + {\rm Re}({\cal F}_{5 4, \alpha} {\cal F}_{5 4, \beta}^*) M_1^2 \right]
 - 2 {\cal C}_{\alpha\beta}^{5 4} {\rm Re}({\cal F}_{4 5, \alpha} {\cal F}_{5 4, \beta}^*) M_2 M_1} \,.\nonumber\\
\end{eqnarray}
Here, ${\cal C}^{\ini \fin}_{\alpha\beta}$ and ${\cal V}^{\ini \fin}_{\alpha\beta}$ were defined in Eq.~\eqref{eq:V_and_C} and the Jarlskog-like parameters are given by 
${\cal J}_{\alpha \beta}^{5 4} = {\rm Im} (R_{\alpha 2} R_{\alpha 1}^* R_{\beta 2}^* R_{\beta 1})$
and
${\cal V}_{\alpha \beta}^{5 4} = {\rm Im} (R_{\alpha 2} R_{\alpha 1}^* R_{\beta 2} R_{\beta 1}^*)$.

\begin{figure}[t!]
    \centering
     \includegraphics[width=0.45\textwidth]{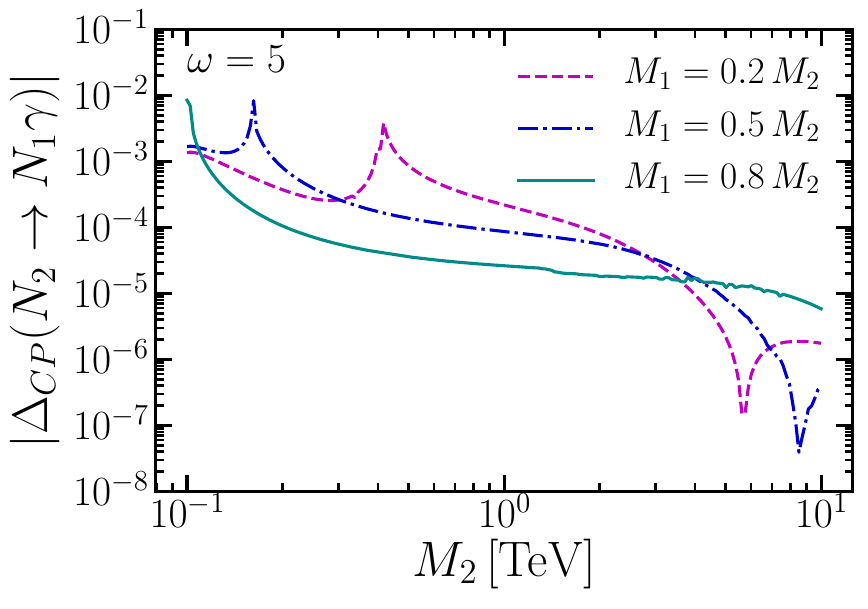}
         \includegraphics[width=0.45\textwidth]{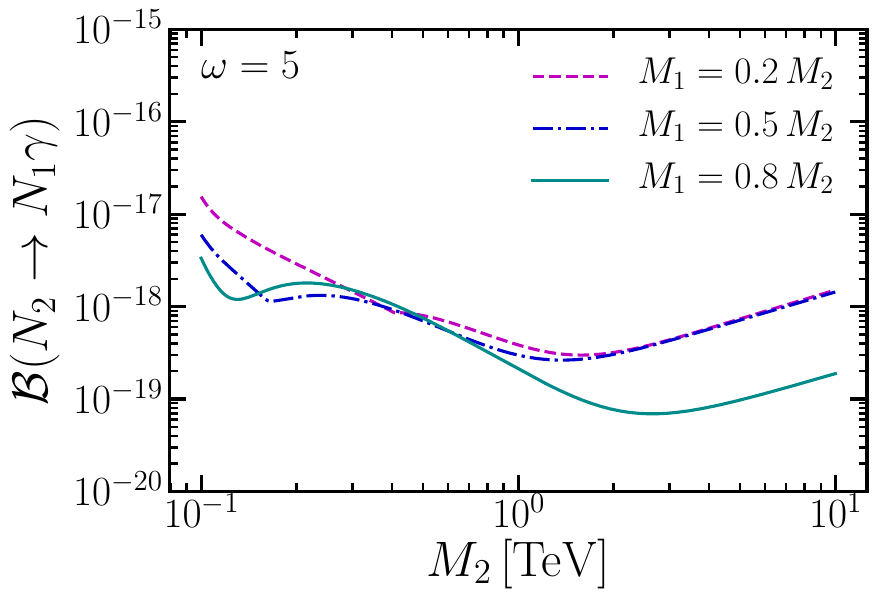}\\
     \includegraphics[width=0.45\textwidth]{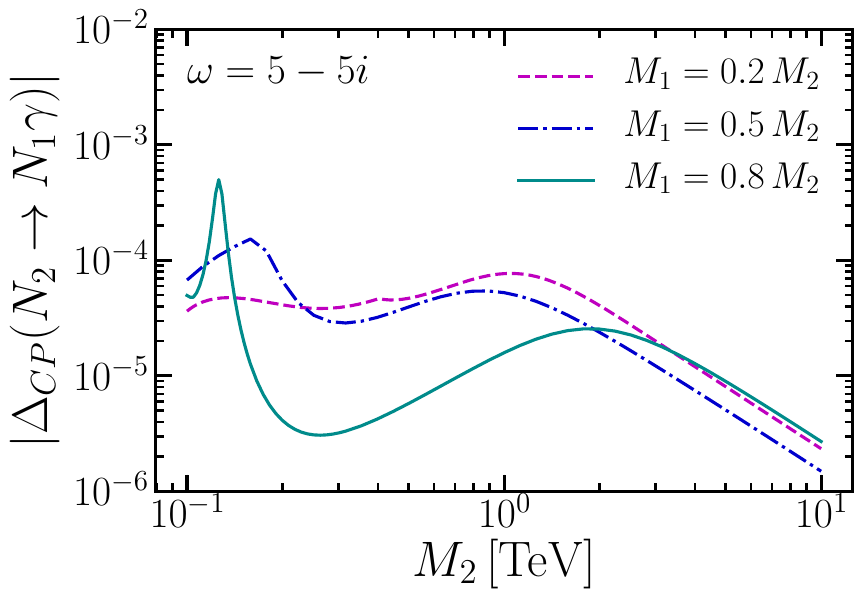}
         \includegraphics[width=0.45\textwidth]{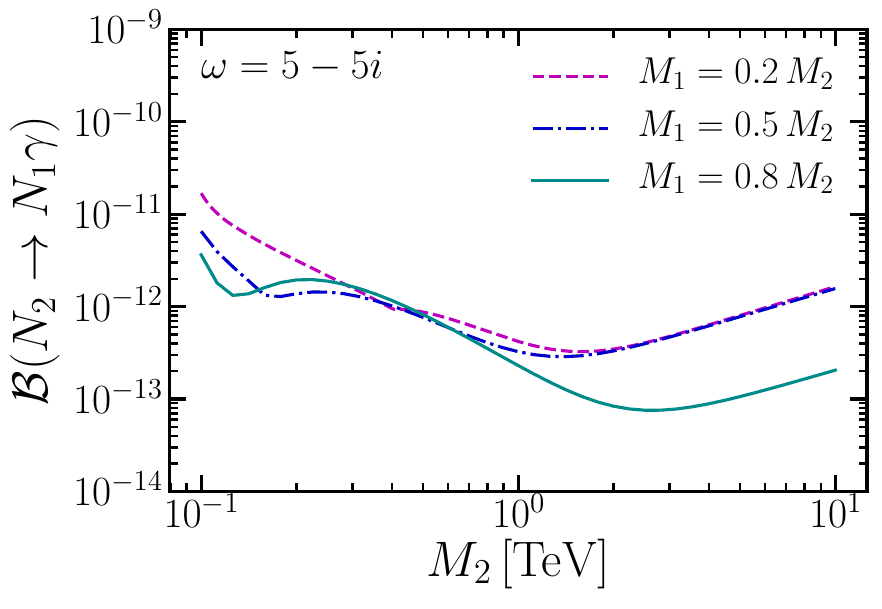}
    \caption{The $CP$ asymmetry (left panel) and branching ratio (right panel) for the radiative decay process $N_2\to N_1\gamma$ as a function of the heavy neutrino mass $M_2$. Four different benchmarks for the lightest right-handed neutrino $M_1=0.2 M_2, 0.5 M_2, 0.8 M_2$ are considered as per the respective plot legends. Values of $\omega$ are fixed at $\omega=5$ (top panel) and $5-5i$ (bottom panel), respectively.  In all cases, we use the best-fit oscillation data as inputs while we set $\zeta=1$ with a Majorana phase $\sigma=\pi/2$.}
    \label{fig:benchmark}
\end{figure}

The behaviour of the $CP$ asymmetry as a function of the right-handed neutrino mass $M_2$ is shown in Fig.~\ref{fig:benchmark}. We can see that the $CP$ asymmetry of this channel is much larger than that in $N \to \nu\gamma$.  In this figure, we vary $M_2$ from 0.1 to 10~TeV and consider three benchmark scenarios where the mass ratio $M_1/M_2$ is fixed to 0.2, 0.5 and 0.8 respectively. In all plots, we fix $\zeta=1$ and the Majorana phase $\sigma=\pi/2$. Therefore, no Majorana-type $CP$ violation is induced. We use the best-fit oscillation data as inputs which include a large $CP$ violating value for $\delta$. In the top panel, we fix $\omega$ to be real, $\omega=5$. Therefore, $\delta$ is the only source of $CP$ violation. 
We note that a large $CP$ asymmetry ratio $|\Delta_{CP}|\sim 10^{-5} \text{-} 10^{-3}$ is easily generated. Peaks of $|\Delta_{CP}|$ are generated due to the enhancement in the log term of ${\rm Im}({\cal F}_{\fin \ini,\alpha})$ around $M_2 \approx m_W$ (cf.~Eq.\eqref{eq:ImF}). Sharp changes refer to cancellations occurring in $\Delta_{CP}$ due to the selected values of inputs. 
In the bottom panel, $\omega = 5-5i$, both $\delta$ and $\omega$ contribute to the $CP$ violation. 
The constraints on $|RR^\dagger|$ from the non-unitarity effect has been included \cite{Fernandez-Martinez:2016lgt}. 

We also show the branching ratio ${\cal B}(N_2\to N_1\gamma) = \Gamma(N_2\to N_1\gamma)/ \Gamma_{N_2}$. In the total decay width $\Gamma_{N_2}$, we include five main decay channels 
$N_2\to \ell^- W^+_{L,T}$, $\nu Z_{L,T}$ and $\nu H$ \cite{Atre:2009rg}. Although the $CP$ asymmetry is large, the branching ratio is suppressed as shown in the right panel of Fig.~\ref{fig:benchmark}, leading to very small $\Delta_{CP}\times {\cal B}$. 
We note that there is particularly interesting phenomenology for $\omega=5-5i$ as the branching ratio is greatly enhanced when assigning an imaginary part to $\omega$. This is because the mixing $R$ is enhanced by $\sin \omega$ and $\cos \omega$, which are both $\sim e^{|\text{Im}[\omega]|}$. One can further increase the branching ratio to be much larger than $ 10^{-13}$ by enlarging the imaginary part of $\omega$, hence the combination $\Delta_{CP} \times {\cal B}$ is also enhanced. 
Another feature of the right panels is that, in spite of the different orders of magnitude, the shape profiles of the curves are almost the same between $\omega=5$ and $5-5i$. This is because the inclusion of an imaginary part for $\omega$ simply changes the size of $R_{\alpha I}$ but rarely changes the correlation between the decay width and right-handed neutrino masses. 

\begin{figure}[t!]
   \centering
     \includegraphics[width=0.45\textwidth]{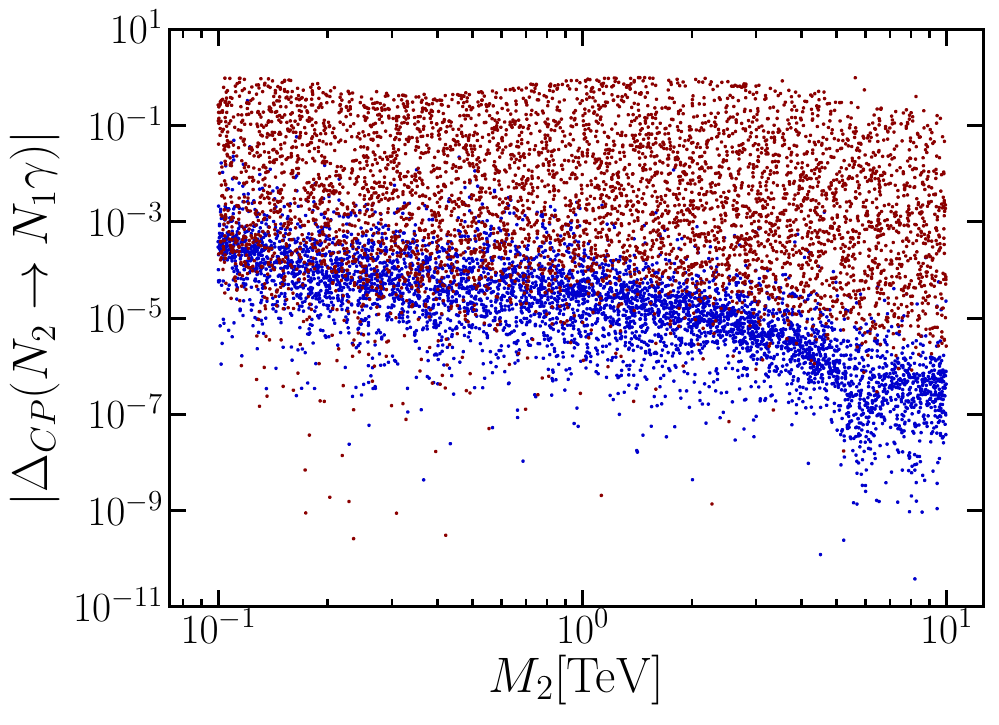}
      \includegraphics[width=0.45\textwidth]{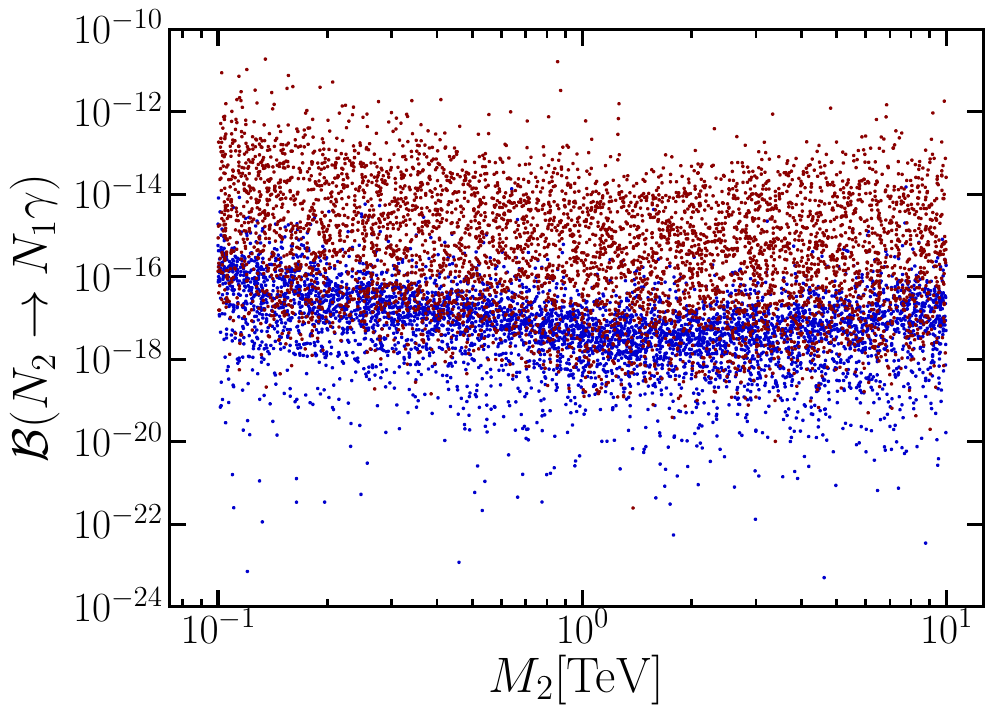}
\caption{The $CP$ asymmetry parameter $\Delta_{CP}$ (left) and branching ratio (right) scanned in the region $M_2$ in $[0.1,10]$~TeV and the ratio $M_1/M_2$ in $[0.1,\;1)$, where both masses are scanned in the logarithmic scale. The red region refers to $\omega=[0,2\pi]+i[-5,5]$ while the blue region is the smaller $\omega=[0,2\pi]$. All oscillation parameters are scanned in the $3\sigma$ ranges, $\omega=[0,2\pi]$ and $\zeta=+1$ are used. The scan performed for the $\zeta=-1$ branch gives the same distribution and is thus omitted.}
\label{fig:TeVNeutrinoRatioScan}
\end{figure}

In Fig. \ref{fig:TeVNeutrinoRatioScan} we show a numerical scan performed for $M_2$ in the same range. We sample $M_2$ logarithmically in the range $[0.1,\,10]$~TeV and the ratio $M_1/M_2$ in the range $[0.1,\,1)$. 
The blue points refer to purely real $\omega$ randomly sampled from $[0, 2\pi)$. In this case, only two of the $CP$ violating phases $\delta$ and $\sigma$ contribute to the $CP$ violation. the $CP$ asymmetry $\Delta_{CP}$ shows a roughly linear correlation  with $M_2^{-1}$. Most points of $\Delta_{CP}$ are located in the regimes $(10^{-3},\;10^{-5})$ for $M_2\simeq 0.1$~TeV, $(10^{-4},\;10^{-6})$ for $M_2\simeq 1$~TeV and $(10^{-5},\;10^{-7})$ for $M_2\simeq 10$~TeV. However, the branching ratio of the decay is tiny, between $( 10^{-20}$, $10^{-15})$, which makes the $CP$ asymmetry unobservable in experiments. For the red points, we allow an imaginary part for $\omega$ as well, namely, ${\rm Im}[\omega] \in [-5, 5]$. A $CP$ asymmetry of order one is then easily achieved. The branching ratio of the radiative decay can maximally reach $\sim10^{-11}$.  We have also checked that the combination $\Delta_{CP} \times {\cal B}$ can maximally reach $4\times 10^{-15}$. 
Note that considering a larger imaginary part of $\omega$ could further enhance the branching ratio and $\Delta_{CP} \times {\cal B}$. However, as this process happens at one loop and there are constraints on the non-unitary effect, the branching ratio is always suppressed by $(16\pi^2)^{-2}|RR^\dag|^2/|RR^\dag|$. By taking $RR^\dag\sim 10^{-3}$, we obtain a branching ratio which maximally reaches $\sim 10^{-7}$ and is therefore challenging to probe in future experiments.


\section{Conclusion\label{sec:conclusion}} 

We study the $CP$ violation in the neutrino electromagnetic dipole moment. A full one-loop calculation of the transition dipole moment is performed in the context of the Standard Model with an arbitrary number of right-handed singlet neutrinos. The $CP$ asymmetry is analytically derived in terms of the leptonic mixing matrix accounting for heavy neutrino mass eigenstates. 
A detailed explanation of how to generate a non-vanishing $CP$ asymmetry in the neutrino transition dipole moment is provided. This requires a threshold condition for the initial neutrino mass being larger than the sum of $W$-boson mass and the charged leptons runnning in the loop and a $CP$ violating phase in the lepton flavour mixing matrix. The threshold condition is necessary to generate a non-zero imaginary part for the loop function. An analytical formulation of this loop integral imaginary component is derived. 
The lepton flavour mixing for vertex contributions has been parametrised in terms of Jarlskog-like parameters. For Majorana particles, the $CP$ asymmetry is identical to the asymmetry of circularly-polarised photons released from the radiative decay. 

The formulation is then applied to a minimal seesaw model where two right-handed neutrinos $N_1$ and $N_2$ are introduced with the mass ordering $M_1 < M_2$. A complete study of $CP$ asymmetry in all radiative decay channels was performed, where the mass range $0.1~{\rm TeV}<M_2<10~{\rm TeV}$ is considered. 
The $CP$ asymmetry in $N_{1,2} \to \nu \gamma$ is very small, maximally reaching $10^{-17}$. In the $N_2 \to N_1 \gamma$ channel, the $CP$ asymmetry is significantly enhanced, with $\Delta_{CP}$ achieving $10^{-5}\text{-}10^{-3}$, even with the Dirac phase $\delta$ being the only source of $CP$ violation. There is a significant correlation between the $CP$ violation in radiative decay and that coming from oscillation experiments. 
We performed a parameter scan of the $CP$ asymmetry with oscillation data in $3\sigma$ ranges taken as inputs and found that the $CP$ asymmetry can maximally reach order one.

\section*{Acknowledgements}

SB is supported by the Australian Research Council (ARC).
MRQ is supported by Consejo Nacional de Ciencia y Tecnologia, Mexico (CONACyT) under grant 440771. YLZ acknowledges the STFC Consolidated Grant ST/L000296/1 and the European Union's Horizon 2020 Research and Innovation programme under Marie Sk\l{}odowska-Curie grant agreements Elusives ITN No.\ 674896 and InvisiblesPlus RISE No.\ 690575.

\end{document}